\PassOptionsToPackage{unicode}{hyperref}
\PassOptionsToPackage{hyphens}{url}
\documentclass[
  10pt,
  letterpaper,
]{article}
\usepackage{xcolor}
\usepackage[margin=1in]{geometry}
\usepackage{amsmath,amssymb}
\usepackage{iftex}
\ifPDFTeX
  \usepackage[T1]{fontenc}
  \usepackage[utf8]{inputenc}
  \usepackage{textcomp} % provide euro and other symbols
\else % if luatex or xetex
  \usepackage{unicode-math} % this also loads fontspec
  \defaultfontfeatures{Scale=MatchLowercase}
  \defaultfontfeatures[\rmfamily]{Ligatures=TeX,Scale=1}
\fi
\usepackage{lmodern}
\ifPDFTeX\else
\fi
\IfFileExists{upquote.sty}{\usepackage{upquote}}{}
\IfFileExists{microtype.sty}{% use microtype if available
  \usepackage[]{microtype}
  \UseMicrotypeSet[protrusion]{basicmath} % disable protrusion for tt fonts
}{}
\makeatletter
\@ifundefined{KOMAClassName}{% if non-KOMA class
  \IfFileExists{parskip.sty}{%
    \usepackage{parskip}
  }{% else
    \setlength{\parindent}{0pt}
    \setlength{\parskip}{6pt plus 2pt minus 1pt}}
}{% if KOMA class
  \KOMAoptions{parskip=half}}
\makeatother
\usepackage{color}
\usepackage{fancyvrb}

\DefineVerbatimEnvironment{Highlighting}{Verbatim}{commandchars=\\\{\}}
\newenvironment{Shaded}{}{}

\newcommand{\NormalTok}[1]{#1}

\usepackage{longtable,booktabs,array}
\usepackage{calc} % for calculating minipage widths
\usepackage{etoolbox}
\makeatletter
\patchcmd\longtable{\par}{\if@noskipsec\mbox{}\fi\par}{}{}
\makeatother
\IfFileExists{footnotehyper.sty}{\usepackage{footnotehyper}}{\usepackage{footnote}}
\makesavenoteenv{longtable}
\usepackage{graphicx}
\makeatletter
\newsavebox\pandoc@box
\newcommand*\pandocbounded[1]{% scales image to fit in text height/width
  \sbox\pandoc@box{#1}%
  \Gscale@div\@tempa{\textheight}{\dimexpr\ht\pandoc@box+\dp\pandoc@box\relax}%
  \Gscale@div\@tempb{\linewidth}{\wd\pandoc@box}%
  \ifdim\@tempb\p@<\@tempa\p@\let\@tempa\@tempb\fi% select the smaller of both
  \ifdim\@tempa\p@<\p@\scalebox{\@tempa}{\usebox\pandoc@box}%
  \else\usebox{\pandoc@box}%
  \fi%
}
\def\fps@figure{htbp}
\makeatother
\NewDocumentCommand\citeproctext{}{}

\makeatletter
 \let\@cite@ofmt\@firstofone
 \def\@biblabel#1{}
 \def\@cite#1#2{{#1\if@tempswa , #2\fi}}
\makeatother
\newlength{\cslhangindent}
\newlength{\csllabelwidth}
\newenvironment{CSLReferences}[2] % #1 hanging-indent, #2 entry-spacing
 {\begin{list}{}{%
  \setlength{\itemindent}{0pt}
  \setlength{\leftmargin}{0pt}
  \setlength{\parsep}{0pt}
  \ifodd #1
   \setlength{\leftmargin}{\cslhangindent}
   \setlength{\itemindent}{-1\cslhangindent}
  \fi
  \setlength{\itemsep}{#2\baselineskip}}}
 {\end{list}}
\usepackage{calc}

\providecommand{\tightlist}{%
  \setlength{\itemsep}{0pt}\setlength{\parskip}{0pt}}
\usepackage{booktabs}
\usepackage{longtable}
\usepackage{graphicx}
\usepackage{microtype}
\usepackage{seqsplit}
\usepackage{bookmark}
\IfFileExists{xurl.sty}{\usepackage{xurl}}{} % add URL line breaks if available
\hypersetup{
  pdftitle={Allostatic Control Systems: Goal Governance in Changing Environments},
  pdfauthor={Thomson D. Nguy},
  hidelinks,
  pdfcreator={LaTeX via pandoc}}

\ifXeTeX

\fi
\ifPDFTeX
\pdftrailerid{}
\pdftrailer{/ID [ <ddc462454c4853e3dcd10d6bcceb7223> <ddc462454c4853e3dcd10d6bcceb7223> ]}
\fi
\ifLuaTeX
\pdfvariable trailerid {[ <ddc462454c4853e3dcd10d6bcceb7223> <ddc462454c4853e3dcd10d6bcceb7223> ]}
\fi

\title{Allostatic Control Systems: Goal Governance in Changing
Environments}
\usepackage{etoolbox}
\makeatletter
\providecommand{\subtitle}[1]{% add subtitle to \maketitle
  \apptocmd{\@title}{\par {\large #1 \par}}{}{}
}
\makeatother
\subtitle{Radiant Institute for Manifold Studies}
\author{Thomson D. Nguy}
\date{}

\begin{document}
\maketitle
\begin{abstract}
Allostatic control systems govern not only how a system pursues a goal,
but whether the goal itself remains appropriate as the environment
changes. This matters whenever a controller can continue to regulate
successfully against a reference that no longer serves the system. We
develop this as a two-timescale problem: a fast loop regulates under the
current reference, while a slow loop governs whether that reference
should move. We then test one candidate mechanism in which slow-loop
movement waits for mature outcome evidence. In a preregistered synthetic
experiment, that mechanism increased decision cost by 1.51\% relative to
otherwise identical bounded adaptation. The central failure was timing:
the evidence-to-effect pathway often could not make correction effective
before the environment changed again. The result motivates a broader
design principle: an allostatic controller must be able to revise an
inappropriate goal faster than serviceability is lost by continuing to
defend it. We therefore treat allostatic control systems as a general
engineering problem of goal governance under changing environments, and
the negative result as one step toward making that problem operational
and falsifiable.
\end{abstract}

\section{Introduction}\label{introduction}

Many engineered systems are organized around a reference condition. A
feedback controller regulates a state toward a target; a monitor
compares measurements with an alarm boundary; a computing platform
allocates resources relative to a service objective; and a classifier
acts on one side or the other of a decision threshold. When the
reference is appropriate, familiar control problems concern sensing,
actuation, delay, noise, model error, constraint handling, and stability
(Astrom and Murray 2008; Rawlings et al. 2026).

There is a second failure mode. A regulator can behave competently
relative to its current reference while the reference itself has become
inappropriate. A threshold suitable for one operating regime may impose
avoidable cost in another; a transient response rule may be mistaken for
a durable change in service posture; or a reference may remain fixed
after the environmental demand it represented has moved. We call this
the \emph{misapplied-homeostat problem}: the system defends a condition
that should itself be reconsidered.

Ordinary control asks how to achieve a goal. Allostatic control asks how
a system should govern the goal itself when a changing environment makes
that goal inappropriate. In this paper, a goal may be represented by a
set point, threshold, reference, service objective, or other defended
condition. The synthetic experiment uses a scalar reference so that one
candidate answer can be isolated and tested; the research question is
broader than that experimental representation.

The deeper cybernetic ancestor is Ashby's ultrastable system. Ashby
coupled primary feedback between a reacting system and its environment
with intermittent, slower feedback that changed the reacting system's
parameters when essential variables passed outside viable limits (Ashby
1960, 80--99). He also briefly identified provisional change of a
sub-goal as an important higher-order possibility while setting it
outside his formal treatment (Ashby 1960, 81). Allostatic Control
Systems place that bracketed question at the center of an engineering
program: when should the condition being defended become an explicit
object of governance? This is an extension of the ultrastability
lineage, not a claim to have discovered second-order adaptation.

Physiology uses \emph{allostasis} for stability through change and,
especially, predictive regulation under anticipated demand (Sterling
2012; Schulkin and Sterling 2019). The term also has direct engineering
precedent. Sanchez-Fibla et al.~used \emph{allostatic control} for a
neurorobotic architecture in which a higher-level regulator coordinated
homeostatic variables by changing desired values and gains
(Sanchez-Fibla et al. 2010). This paper therefore does not coin the term
or introduce the generic idea of an upper layer changing what a lower
layer defends. Building on that lineage, we use \textbf{Allostatic
Control Systems} as a broad engineering program for systems that govern
what their fast regulators defend as environments and demands change.
Such systems may use anticipatory context, higher-level drive
arbitration, outcome evidence, or combinations of those mechanisms. The
present paper develops and tests one narrower, evidence-governed
specialization: a fast loop acts under the current reference, while a
slower loop decides whether delayed, cost-bearing outcome evidence
warrants moving that reference. The aim is to make the broader program
operational, falsifiable, and useful across engineered domains, not to
claim biological equivalence or nomenclatural priority. Adaptive,
supervisory, hierarchical, economic model predictive,
reference-governor, and change-detection methods already contain
neighboring or overlapping machinery (Astrom and Wittenmark 1995;
Ramadge and Wonham 1987; Garone et al. 2017; Amrit et al. 2011; Gama et
al. 2014).

The proposed distinction is evidentiary. Reference movement should be
justified by evidence that the present reference failed to serve demand
under a declared cost, not by raw frequency, usage, or clock time alone.
The slow loop therefore operates in \emph{evidence time}: it waits for
outcomes relevant to the set-point question, rejects stale evidence, and
limits how evidence may be reused. This added discipline is only useful
if it improves outcomes or constrains harm relative to a controller with
the same adaptive mechanics but without those evidence governance gates.

To make this architecture falsifiable rather than merely taxonomic, the
paper subjects one evidence-governance interpretation to a direct
technical test. Its contribution is not the phrase \emph{allostatic
control}, changeable set points, dynamic gains, or layered regulation.
It turns one evidence-governance interpretation into a reproducible
research program with an explicit comparator and a result that can fail.
Specifically, the paper contributes:

\begin{enumerate}
\def\labelenumi{\arabic{enumi}.}
\tightlist
\item
  an operational two-timescale reference-governance model with delayed
  evidence, bounded direction-specific movement, and explicit
  admissibility constraints;
\item
  a matched ablation that isolates evidence maturity, staleness,
  ordering, and consumption from adaptation itself;
\item
  a preregistered synthetic experiment with fixed, naive adaptive,
  periodic supervisory, matched ungated, and allostatic policies,
  development-only parameter selection, fresh confirmatory streams, and
  immutable execution receipts; and
\item
  a negative confirmatory result that separates available oracle
  headroom from the tested controller's inability to exploit it, and
  identifies evidence latency and gated information loss as candidate
  failure conditions.
\end{enumerate}

The central hypothesis did not survive its registered test. In the
tested environment, mature-cohort governance is not an empirically
supported cost-improving design method. The broader architectural
distinction remains useful as a taxonomy and as a way to formulate
experiments; the negative result applies to this mechanism and declared
regime, not to every possible form of reference governance.

\section{Definition, Scope, and Screening
Criteria}\label{definition-scope-and-screening-criteria}

Let \(t\) index fast decisions and \(j\) index slow reference versions.
The defended reference \(r_j\) belongs to a declared admissible set
\(\mathcal{R}\). A fast policy acts on observation \(z_t\) under the
currently active reference,

\[
a_t = \pi_{\mathrm{fast}}(z_t,r_j).
\]

The slow policy receives delayed operating evidence and maps an eligible
evidence object \(E_j\) and current reference to a successor reference,

\[
r_{j+1} = \pi_{\mathrm{slow}}(r_j,E_j), \qquad r_{j+1}\in\mathcal{R}.
\]

At the broader program level, an engineered allostatic control system
makes the goal or defended condition an explicit object of governance
rather than assuming it should remain fixed. The mechanism that warrants
movement is domain-dependent. It may be predictive, contextual,
drive-based, outcome-based, or hybrid.

The architecture satisfies the evidence-governed specialization tested
here when: (i) the reference is an explicit governance object rather
than an incidental tuned parameter; (ii) slow updates are justified by
outcome evidence bearing on reference mismatch; (iii) evidence
eligibility is distinct from the fast decision clock; and (iv) movement
is constrained by declared admissibility and rate limits.

This specialization is deliberately narrower than ``any changing
threshold.'' It also does not make allostatic control necessary whenever
an environment is non-stationary. A candidate application should satisfy
five operational screening questions:

{\def\LTcaptype{none} % do not increment counter
\begin{longtable}[]{@{}
  >{\raggedright\arraybackslash}p{(\linewidth - 4\tabcolsep) * \real{0.3333}}
  >{\raggedright\arraybackslash}p{(\linewidth - 4\tabcolsep) * \real{0.3333}}
  >{\raggedright\arraybackslash}p{(\linewidth - 4\tabcolsep) * \real{0.3333}}@{}}
\toprule\noalign{}
\begin{minipage}[b]{\linewidth}\raggedright
Criterion
\end{minipage} & \begin{minipage}[b]{\linewidth}\raggedright
Observable check
\end{minipage} & \begin{minipage}[b]{\linewidth}\raggedright
Counterexample
\end{minipage} \\
\midrule\noalign{}
\endhead
\bottomrule\noalign{}
\endlastfoot
Demand-dependent reference & Domain evidence identifies contexts in
which a different target, threshold, or posture is appropriate before
the fast action. & The same reference remains appropriate and only
control effort changes. \\
Non-stationary demand & Regime shifts, drift, seasonality, or strategic
response alter what should be defended. & Variation is noise around a
stable operating regime. \\
Asymmetric or state-dependent cost & Errors on opposite sides of the
reference have materially different consequences. & Deviations are
approximately symmetric and reversible. \\
Partial anticipatability & Context or leading evidence can be observed
before mismatch is fully realized. & Evidence arrives only after
irrecoverable failure. \\
Observable failed serviceability & A cost-bearing outcome can be
attributed to the current reference and compared with alternatives. &
The only signal is frequency, popularity, or operator preference. \\
\end{longtable}
}

Failure of these checks is informative. If the reference need not move
independently of fast action, existing feedback or adaptive control is
sufficient. If outcome evidence cannot be attributed to reference
mismatch, autonomous movement is not warranted. If evidence matures more
slowly than regimes persist, a maturity gate can make adaptation
systematically late.

\subsection{Serviceability evidence that can move a
goal}\label{serviceability-evidence-that-can-move-a-goal}

A slow loop should not be licensed by every change in a metric.
Frequency, usage, alarm count, load, detection volume, or operator
preference may reflect exposure, noise, policy, habit, or gaming rather
than a goal that has become inappropriate. We use \emph{scarce-resource
evidence} for evidence that maintaining the current reference consumed a
constrained capacity or imposed a meaningful cost in a way that is
diagnostic of reference mismatch. Depending on the domain, the scarce
resource may be attention, response capacity, recovery time, compute,
safety margin, missed opportunity, or another limited capacity the
system is meant to protect.

Evidence capable of moving a defended goal should pass five checks:

\begin{enumerate}
\def\labelenumi{\arabic{enumi}.}
\tightlist
\item
  \textbf{Demand-revealing:} it bears on the environment the system is
  meant to serve, not merely on what the current policy exposed or
  counted.
\item
  \textbf{Reference-specific:} it bears on whether the defended goal was
  wrong, not merely on whether the system experienced a bad outcome.
\item
  \textbf{Cost-bearing:} it records a constrained resource, burden, or
  opportunity cost rather than a signal that can be produced freely.
\item
  \textbf{Differentially attributable:} it helps distinguish reference
  mismatch from failures of sensing, actuation, modeling, or execution.
\item
  \textbf{Policy-aware:} it accounts for the possibility that the active
  policy shaped, censored, or made the evidence easy to manipulate.
\end{enumerate}

The practical test is counterfactual substitutability: would an
admissible nearby reference have served the observed demand better under
the declared cost? If no plausible alternative would have reduced the
burden, a costly outcome does not by itself show that the goal was
wrong. In the experiment below, empirical decision risk is one
deliberately small instantiation of this idea. It does not make generic
cost sufficient evidence for allostatic movement in other domains.

\section{Relation to Existing Control
Methods}\label{relation-to-existing-control-methods}

Ashby's ultrastability is the foundational two-feedback antecedent.
Primary feedback regulates the acting system within its current
organization; slower feedback changes that organization when essential
variables leave their acceptable limits (Ashby 1960, chap. 7). In
Ashby's formal construction, the limits on essential variables supply
the adaptation criterion. The present program asks a different but
adjacent question: when environmental change makes a defended reference
or objective inappropriate, what evidence licenses changing that
defended condition itself?

The direct terminological and architectural precedent is bio-inspired
allostatic control. Sanchez-Fibla et al.~modeled an allostatic regulator
above homeostatic variables, dynamically altering desired values and
gains in comparative rodent-robot experiments (Sanchez-Fibla et al.
2010). The present paper does not claim priority over that architecture.
Its narrower question is whether delayed outcome evidence provides a
useful warrant for reference movement, and its experiment tests that
warrant against a mechanically matched ungated policy.

Within classical control, the nearest structural boundary is supervisory
or hierarchical control, not fixed feedback. Supervisory control can
select modes, policies, or references for a lower-level controller
(Ramadge and Wonham 1987). Adaptive control can update parameters or
models from observed behavior (Astrom and Wittenmark 1995). Model
predictive and economic model predictive control can optimize
trajectories, constraints, and economic objectives over a horizon (Mayne
et al. 2000; Amrit et al. 2011; Rawlings et al. 2012). Concept drift and
change-detection methods can identify distributional change (Gama et al.
2014). Reference and command governors explicitly modify reference
inputs to preserve constraints in pre-stabilized systems (Garone et al.
2017). Learning-based MPC treats learned performance improvement and
constraint or safety certification as distinct design obligations
(Hewing et al. 2020). The present construction provides no comparable
safety certificate or stability guarantee; projection onto an admissible
interval is only an implementation constraint. Two-timescale stochastic
approximation provides a mathematical language for coupled processes
evolving at different rates (Borkar 1997).

Allostatic control does not replace those methods. It poses a design
question that can be implemented through them: what evidence licenses
movement of the condition being defended? The answer may be embedded in
a supervisory policy, an adaptive law, a reference governor, or a
predictive controller. The contribution is therefore a discipline and an
experimental boundary, not priority over changing references or layered
control.

The novelty claim is correspondingly specific. This paper does not claim
to originate the term \emph{allostatic control system}, a new class of
stabilizing controller, higher-level coordination of homeostatic loops,
or dynamic adjustment of desired values or gains. It makes one
delayed-evidence reference-governance hypothesis experimentally
separable from adaptation by pairing the proposed policy with an ungated
controller that shares its local law and tuning space. That matched
contrast, together with preregistered failure gates and preservation of
the negative result, is the technical contribution evaluated here.

{\def\LTcaptype{none} % do not increment counter
\begin{longtable}[]{@{}
  >{\raggedright\arraybackslash}p{(\linewidth - 6\tabcolsep) * \real{0.2500}}
  >{\raggedright\arraybackslash}p{(\linewidth - 6\tabcolsep) * \real{0.2500}}
  >{\raggedright\arraybackslash}p{(\linewidth - 6\tabcolsep) * \real{0.2500}}
  >{\raggedright\arraybackslash}p{(\linewidth - 6\tabcolsep) * \real{0.2500}}@{}}
\toprule\noalign{}
\begin{minipage}[b]{\linewidth}\raggedright
Method
\end{minipage} & \begin{minipage}[b]{\linewidth}\raggedright
Primary object changed
\end{minipage} & \begin{minipage}[b]{\linewidth}\raggedright
Typical update basis
\end{minipage} & \begin{minipage}[b]{\linewidth}\raggedright
Boundary with this work
\end{minipage} \\
\midrule\noalign{}
\endhead
\bottomrule\noalign{}
\endlastfoot
Fixed feedback & Action under a specified reference & Instantaneous or
filtered error & Baseline when the reference remains appropriate. \\
Adaptive control / self tuning & Model or controller parameters &
Identification or performance error & Becomes allostatic here only when
the defended reference is an explicit governed object. \\
Gain scheduling & Controller configuration & Operating point or regime
indicator & Changes behavior across regimes; need not govern the warrant
for reference movement. \\
Supervisory / hierarchical control & Mode, policy, goal, or reference &
Higher-level logic & Closest structural neighbor; can implement the
proposed evidence discipline. \\
MPC / economic MPC & Planned action, trajectory, or objective & Forecast
model, constraints, cost & Compatible architecture; this work does not
claim novelty over MPC. \\
Change detection / adaptive thresholding & Detector state or threshold &
Distributional evidence or local error & Detecting change does not by
itself specify whether a defended reference should move. \\
Ashbian ultrastability & Parameters governing the reacting system &
Essential variables leaving viable limits & Foundational second-order
adaptation architecture; the formal viability limits remain the
criterion rather than a governed reference. \\
Bio-inspired allostatic control & Desired values, gains, or drive
priorities & Higher-level regulation of homeostatic variables & Direct
terminological and architectural precedent for changing what lower loops
defend. \\
Evidence-governed allostatic adaptation (this study) & Defended
reference & Delayed cost-bearing mismatch evidence & A narrower
specialization whose value must be tested against a mechanically matched
ungated policy. \\
\end{longtable}
}

This boundary determines the primary falsification criterion. Beating a
fixed reference would only show that adaptation helped. The distinctive
claim requires mature-evidence governance to beat an adaptive comparator
sharing the same local update mechanics and tuning space.

\section{Formal Reference-Governance
Model}\label{formal-reference-governance-model}

\subsection{Evidence time and a first-order viability
condition}\label{evidence-time-and-a-first-order-viability-condition}

Evidence time measures progress through outcome-bearing events rather
than clock time alone. Its units may be resolved decisions, patient
episodes, response cycles, workload regimes, recovery intervals, or
other opportunities in which the current reference could have succeeded
or failed. Let \(T_r\) denote recovery time after a reference or policy
change, \(T_m\) the time required for evidence to mature, and \(T_e\)
the persistence time of the relevant environmental regime. A first-order
viability condition is

\[
T_r \leq T_m < T_e.
\]

The left inequality guards against mistaking the system's own recovery
transient for changed demand. The right guards against acting on
evidence from a regime that no longer exists. This is a screening
heuristic, not a stability theorem, and it is not sufficient: outcome
reveal, eligibility, slow-loop use, bounded movement, and material
downstream effect can consume additional time after nominal maturity.
The experiment below tests one concrete maturity rule; the
interpretation returns to the stronger end-to-end timing problem exposed
by its failure.

\subsection{Delayed evidence}\label{delayed-evidence}

For the synthetic experiment, \(z_t\in[0,1]\) is an anonymous score and
\(y_t\in\{0,1\}\) is a latent label indicating whether a distinct
synthetic referent is warranted. The fast action is

\[
a_t(r_j)=\mathbf{1}\{z_t\ge r_j\}.
\]

Labels are not immediately available. If an outcome is revealable, it
arrives after a random delay as the immutable record \((t,z_t,y_t)\).
Unrevealed labels are unavailable to every policy. Decisions are
assigned to non-overlapping evidence cohorts. At a slow checkpoint, a
cohort is eligible only if it has reached a minimum resolved count and
fraction, is not stale, and has not previously been consumed by the
allostatic policy.

For resolved records in cohort \(E\), define empirical decision risk at
candidate reference \(r\),

\[
J_E(r)=\frac{1}{n_E}\sum_{t\in E}\left[
c_{\mathrm{FI}}\mathbf{1}\{z_t\ge r,y_t=0\}
+c_{\mathrm{SD}}\mathbf{1}\{z_t<r,y_t=1\}\right],
\]

where \(c_{\mathrm{FI}}\) is the cost of false instantiation,
\(c_{\mathrm{SD}}\) is the cost of a suppressed distinction, and \(n_E\)
is the number of resolved outcomes. Revealed pairs permit counterfactual
evaluation of nearby references without revealing any missing label.

\subsection{Local signal and bounded asymmetric
update}\label{local-signal-and-bounded-asymmetric-update}

For perturbation \(\epsilon>0\) and admissible interval
\(\mathcal{R}=[r_{\min},r_{\max}]\), define

\[
e_j = J_{E_j}\!\left(\max(r_{\min},r_j-\epsilon)\right)
-J_{E_j}\!\left(\min(r_{\max},r_j+\epsilon)\right).
\]

A positive signal indicates lower local empirical risk under a stricter
reference; a negative signal indicates lower risk under a looser
reference. With deadband \(\delta\ge0\), directional gains
\(g_+,g_->0\), and step caps \(\eta_+,\eta_->0\), the slow update is

\[
\Delta_j =
\begin{cases}
0, & |e_j|\le\delta,\\
\min\{\eta_+,g_+(e_j-\delta)\}, & e_j>\delta,\\
-\min\{\eta_-,g_-(-e_j-\delta)\}, & e_j<-\delta,
\end{cases}
\]

\[
r_{j+1}=\Pi_{\mathcal R}(r_j+\Delta_j).
\]

For the allostatic policy, absent, immature, stale, or previously
consumed evidence yields \(r_{j+1}=r_j\). Eligible cohorts are
considered oldest first and consumed at most once even if the deadband
produces a hold. These choices are implementation hypotheses, not
consequences of the allostatic definition.

\subsection{Construction invariants}\label{construction-invariants}

The following statements are implementation sanity checks that follow
immediately from the update construction. They are not offered as
theorem-level control contributions or plant-level claims; the
analytical burden of this study is the matched empirical mechanism test
below.

\textbf{Invariant 1 (admissibility).} If \(r_j\in\mathcal R\), then
\(r_{j+1}\in\mathcal R\).

\emph{Proof.} \(\Pi_{\mathcal R}\) maps every scalar argument into
\(\mathcal R\). \(\square\)

\textbf{Invariant 2 (bounded movement).} Each slow update satisfies

\[
|r_{j+1}-r_j|\le \max(\eta_+,\eta_-).
\]

\emph{Proof.} The unprojected increment is bounded by the corresponding
directional cap, and scalar projection onto an interval is
non-expansive. \(\square\)

\textbf{Invariant 3 (hold and direction).} Ineligible evidence or a
signal in the deadband cannot change the reference. An eligible positive
signal cannot loosen it, and an eligible negative signal cannot tighten
it.

These properties establish interval membership, a one-update movement
limit, and sign consistency only. They do not establish closed-loop
stability, convergence, optimality, regret, robustness, or safety.

\subsection{Matched ungated ablation}\label{matched-ungated-ablation}

The matched comparator uses the same \(J_E\), \(\epsilon\), deadband,
gains, step caps, projection, checkpoint schedule, initial reference,
and joint tuning grid. It differs only in evidence governance: at each
checkpoint it uses the most recent fixed-size window of revealed events,
allows reuse across checkpoints, and applies no cohort-maturity,
staleness, or one-consumption gate. The allostatic-minus-matched
contrast therefore estimates the effect of the tested
evidence-governance rules, not the effect of bounded adaptation in
general.

The model is summarized in Algorithm 1.

\begin{Shaded}
\begin{Highlighting}[]
\NormalTok{Initialize admissible reference r and version j = 0}
\NormalTok{For each fast decision t:}
\NormalTok{    observe anonymous score z\_t}
\NormalTok{    act under the current reference r}
\NormalTok{    attach a delayed synthetic label only when its reveal event occurs}
\NormalTok{At each slow checkpoint:}
\NormalTok{    retire stale cohorts}
\NormalTok{    choose the oldest mature, non{-}stale, unused cohort E}
\NormalTok{    if no E exists, hold r}
\NormalTok{    otherwise compute the local revealed{-}risk signal}
\NormalTok{    apply the deadbanded, direction{-}specific, capped update}
\NormalTok{    project the successor reference into the admissible interval}
\NormalTok{    mark E consumed and increment the policy version}
\end{Highlighting}
\end{Shaded}

\section{Falsifiable Design Criteria and Failure
Modes}\label{falsifiable-design-criteria-and-failure-modes}

The formal model makes several failure modes observable rather than
rhetorical.

\begin{enumerate}
\def\labelenumi{\arabic{enumi}.}
\tightlist
\item
  \textbf{Noise chasing.} A small or absent deadband can convert
  sampling variation into repeated movement.
\item
  \textbf{Evidence latency.} Maturity requirements can delay action
  until the relevant regime has ended.
\item
  \textbf{Stale-evidence starvation.} Strict maturity under incomplete
  reveal can cause cohorts to expire before use.
\item
  \textbf{Policy-shaped evidence.} Observed outcomes can depend on the
  active policy, weakening attribution from cost to reference mismatch.
\item
  \textbf{Shallow cost surface.} Meaningful oracle movement need not
  produce a meaningful aggregate cost advantage over a fixed compromise
  reference.
\item
  \textbf{Governance overhead.} Ordering and one-consumption rules can
  discard useful recent information without compensating benefit.
\item
  \textbf{Boundary saturation.} Repeated one-sided evidence can drive a
  reference to its admissible limit, where projection hides further
  pressure.
\end{enumerate}

These are not merely implementation cautions. Each supplies a reason the
distinctive allostatic mechanism may add no engineering value. The
registered experiment was designed so that the matched ablation,
stationary control, rapid-shift boundary, and no-signal control could
expose them.

\section{Preregistered Synthetic
Experiment}\label{preregistered-synthetic-experiment}

\subsection{Research question and
generator}\label{research-question-and-generator}

The preregistered question was: under delayed and incomplete outcomes
and bidirectional prevalence changes, does mature-cohort evidence
governance improve a bounded adaptive scalar reference relative to a
matched ungated policy and periodic empirical-risk supervision?

Each run contained 4,000 decisions. In the primary scenario, latent
labels followed \(Y_t\sim\mathrm{Bernoulli}(p_q)\) over four consecutive
1,000-decision segments with \(p_q=(0.35,0.15,0.65,0.35)\). Conditional
scores were

\[
Z_t\mid Y_t=1\sim\mathrm{Beta}(8,3),\qquad
Z_t\mid Y_t=0\sim\mathrm{Beta}(3,8).
\]

Each label was revealable independently with probability 0.75.
Revealable outcomes arrived after a geometrically distributed delay with
parameter 0.02, capped at 600 decisions. A reveal tail through boundary
4,600 ensured that all revealable outcomes could be scored while
preventing controller updates after decision 4,000. False instantiation
cost 1 and suppressed distinction cost 2. The reference interval was
\([0.25,0.75]\).

Three additional scenarios were frozen. A stationary safety control held
\(p=0.35\). A rapid boundary alternated \(p=0.15\) and \(p=0.65\) every
100 decisions. In the independent-label null, scores followed
\(\mathrm{Beta}(5,5)\) independently of labels, while label prevalence
followed the primary schedule.

For internal adaptation-delay measurement, the generator admits a
closed-form Bayes threshold. Because the beta normalizing constants are
equal,

\[
\frac{f_1(z)}{f_0(z)}=\left(\frac{z}{1-z}\right)^5,
\]

and the cost-optimal threshold at prevalence \(p\) is

\[
r^*(p)=\frac{K(p)^{1/5}}{1+K(p)^{1/5}},\qquad
K(p)=\frac{c_{\mathrm{FI}}(1-p)}{c_{\mathrm{SD}}p}.
\]

This gives approximate oracle references 0.552, 0.496, and 0.435 at
prevalence 0.15, 0.35, and 0.65. Controllers never received the oracle.

The same generator makes a headroom check possible. If \(F_{a,b}\) is
the beta cumulative distribution function, expected decision cost at
reference \(r\) and prevalence \(p\) is

\[
R(r;p)=(1-p)\left[1-F_{3,8}(r)\right]+2pF_{8,3}(r).
\]

Across the four equal-length primary segments, unconstrained Bayes
placement yields expected mean cost 0.067515, compared with 0.074799 for
the selected fixed reference of 0.49. Relative to the observed
matched-ungated mean of 0.074758, the difference is 9.69\%. This
calculation allows the reference to jump at each segment boundary; it
establishes reference-placement headroom in the cost surface, not
attainable performance for a movement-limited controller.

A second post-result diagnostic imposed the registered 50-decision
checkpoint schedule and the selected 0.01 directional step caps.
Starting at 0.49, a clairvoyant tracker that moved directly toward the
current segment's Bayes reference achieved expected mean cost 0.071007
with zero detection lag, or 5.02\% improvement relative to matched
ungated. With imposed lags of 50, 100, 150, and 200 decisions, the
corresponding improvements were 3.33\%, 1.63\%, -0.06\%, and -1.75\%.
Thus one capped reference trajectory with immediate regime knowledge
clears the 3\% gate, but the margin decays quickly with lag. These
diagnostic trajectories neither give oracle information to a tested
controller nor establish reachability or impossibility for a realizable
evidence-governed policy. They do not alter the registered result.

\subsection{Policies and parameter
selection}\label{policies-and-parameter-selection}

All five policies received identical generated decisions, reveal
schedules, initial reference, and observable evidence.

\begin{itemize}
\tightlist
\item
  \textbf{Fixed} held a development-selected reference.
\item
  \textbf{Naive adaptive} updated from newly revealed events in the
  latest checkpoint interval using a symmetric bounded local signal,
  without maturity or staleness screening.
\item
  \textbf{Periodic supervisory} selected the empirical-risk-minimizing
  point on a 0.01 reference grid at fixed checkpoint periods using a
  trailing revealed-event window.
\item
  \textbf{Matched ungated} used the same local law and tuning space as
  allostatic control but acted on the most recent revealed events
  without evidence-governance gates.
\item
  \textbf{Allostatic} used the oldest mature, non-stale, unconsumed
  cohort.
\end{itemize}

Slow checkpoints occurred every 50 decisions, for 79 opportunities.
Allostatic cohorts also contained 50 decisions, required at least 25
resolved labels and a 0.70 resolved fraction, and became stale 300
decisions after cohort end.

Parameters were selected on 50 development runs for each of the primary
and stationary scenarios. The search comprised 21,200 registered
policy-configuration evaluations. Mean decision cost was the objective
and total reference variation was a tie break. The allostatic and
matched policies shared one selected local-law configuration. The
selected initial reference was 0.49. The allostatic/matched parameters
were \(\epsilon=0.03\), \(\delta=0.02\), \(g_+=g_-=0.10\), and
\(\eta_+=\eta_-=0.01\). Development results were not used to alter the
generator, hypotheses, parameter grids, or confirmatory protocol.

Joint selection of the allostatic and matched configuration was part of
the estimand, not a claim that either policy was globally optimized. It
prevents a tuning advantage from being mistaken for an effect of
evidence governance. H1 therefore estimates the incremental effect of
the maturity, staleness, ordering, and one-consumption bundle
conditional on the shared development-selected local law.

Confirmatory independence was therefore at the random-stream level, not
at the generator-family level. Development selection and confirmation
used fresh streams but the same declared primary and stationary scenario
classes. Effect sizes, the selected fixed reference, and the selected
adaptive law are conditional on that generator family; no out-of-family
performance claim is made.

\subsection{Hypotheses and estimands}\label{hypotheses-and-estimands}

One hundred fresh paired runs per scenario were generated after
parameter freeze.

\textbf{H1 (mechanism).} On the primary shift, allostatic control must
improve mean combined cost relative to matched ungated adaptation. The
registered decision rule used a one-sided 95\% percentile-bootstrap
lower bound on relative improvement, which had to exceed zero, and a
point estimate of at least 3\%. The analysis also tabulated the
corresponding one-sided upper summary to show the observed effect's
direction; that upper value was not a second H1 gate.

\textbf{H3 (stationary safety).} On the stationary control, allostatic
cost excess had to remain below 2\% relative to both fixed and periodic
policies, and the one-sided upper bound on median maximum reference
drift had to be at most 0.03.

\textbf{H2 (conditional refinement).} Only if the central claim passed,
allostatic control had to be cost-noninferior to periodic supervision
and reduce total reference variation by at least the registered margin.
The hierarchy forbade H3 or H2 components from becoming independent
success claims after an H1 failure.

Run-index clusters were the unit of paired inference. Registered
bootstrap intervals used 10,000 resamples; paired sign-flip tests used
10,000 permutations. Descriptive pairwise tests were Holm adjusted
within scenario and metric. No run was excluded or replaced.

\subsection{Execution custody and
reproducibility}\label{execution-custody-and-reproducibility}

The preregistration, source tree, runtime lock, and selected parameters
were hashed before a confirmatory authorization salt existed. A
KMS-signed receipt bound those hashes to a single atomic execution
claim. The run produced 2,000 primary policy results and 3,500
sensitivity rows. Raw results, telemetry, and the execution manifest
were written as separate versioned objects under governance retention.
The analysis was independently replayed and produced the same SHA-256
digest.

Completeness checks found 2,000 unique scenario-run-policy keys, 400
paired streams, 79 slow checkpoints per run, no nonfinite metrics, and
no reference-bound violations. The execution identifier was
\seqsplit{220c98da7931c43b957c8bd86ccb83ea}; the frozen preregistration
digest was
\seqsplit{b66740c8cc986bf9876e3e10ada043ec5e390bc855fbb6c2614a1709e53008d0};
and the independently replayed analysis digest was
\seqsplit{0e103002f718ba3e6a8eb4103369e27add629346b323a19086ab3f862a67dd92}.

\section{Results}\label{results}

\subsection{Central mechanism test}\label{central-mechanism-test}

H1 failed both registered gates. Mean primary cost was 0.0758875 for
allostatic control and 0.0747575 for matched ungated adaptation.
Relative improvement,
\((C_{\mathrm{matched}}-C_{\mathrm{allostatic}})/C_{\mathrm{matched}}\),
was -0.01512. Its registered one-sided 95\% lower bound was -0.01906;
the corresponding upper summary was -0.01141. Both lay on the harmful
side of zero, and the point estimate was far below the registered 3\%
practical gate. This was not an ambiguous null: the tested
mature-evidence policy bundle increased cost by 1.51\% relative to its
matched ablation.

The headroom calculation separates failure to win from active harm.
Unconstrained Bayes placement was 9.69\% below the observed matched
mean, ruling out a flat cost surface but overstating what a
movement-limited controller could attain. A clairvoyant trajectory
obeying the registered checkpoint schedule and 0.01 step cap achieved
5.02\% improvement with zero detection lag; at 100 decisions of imposed
lag, its improvement fell to 1.63\%. The calculation therefore
establishes that an immediate-detection capped trajectory could clear
3\%, not that a realizable delayed-evidence policy could. Independently,
the allostatic bundle performed 1.51\% worse than its mechanically
matched comparator. That paired comparison is the registered empirical
finding.

{\def\LTcaptype{none} % do not increment counter
\begin{longtable}[]{@{}
  >{\raggedright\arraybackslash}p{(\linewidth - 10\tabcolsep) * \real{0.1304}}
  >{\raggedleft\arraybackslash}p{(\linewidth - 10\tabcolsep) * \real{0.1739}}
  >{\raggedleft\arraybackslash}p{(\linewidth - 10\tabcolsep) * \real{0.1739}}
  >{\raggedleft\arraybackslash}p{(\linewidth - 10\tabcolsep) * \real{0.1739}}
  >{\raggedleft\arraybackslash}p{(\linewidth - 10\tabcolsep) * \real{0.1739}}
  >{\raggedleft\arraybackslash}p{(\linewidth - 10\tabcolsep) * \real{0.1739}}@{}}
\toprule\noalign{}
\begin{minipage}[b]{\linewidth}\raggedright
Primary policy
\end{minipage} & \begin{minipage}[b]{\linewidth}\raggedleft
Mean (SD) cost
\end{minipage} & \begin{minipage}[b]{\linewidth}\raggedleft
Median {[}IQR{]} cost
\end{minipage} & \begin{minipage}[b]{\linewidth}\raggedleft
Mean total variation
\end{minipage} & \begin{minipage}[b]{\linewidth}\raggedleft
Mean holds
\end{minipage} & \begin{minipage}[b]{\linewidth}\raggedleft
Mean stale retirements
\end{minipage} \\
\midrule\noalign{}
\endhead
\bottomrule\noalign{}
\endlastfoot
Fixed & 0.074785 (0.005112) & 0.074500 {[}0.006250{]} & 0.000000 & 79.00
& 0.00 \\
Naive adaptive & 0.074558 (0.005207) & 0.074375 {[}0.006250{]} &
0.420496 & 13.26 & 0.00 \\
Periodic supervisory & 0.077588 (0.005967) & 0.077375 {[}0.008000{]} &
0.534300 & 62.79 & 0.00 \\
Matched ungated & 0.074758 (0.005335) & 0.074125 {[}0.006750{]} &
0.142655 & 40.47 & 0.00 \\
Allostatic & 0.075888 (0.005322) & 0.075500 {[}0.006250{]} & 0.149590 &
31.06 & 11.63 \\
\end{longtable}
}

All 100 run-level outcomes per policy were retained. Appendix B reports
all-policy summaries for every registered scenario, directional segment
costs, and the complete sensitivity grid.

\begin{figure}
\centering
\includegraphics[width=1\linewidth,height=\textheight,keepaspectratio,alt={Distribution summaries for decision cost and total reference variation across the 100 paired confirmatory runs. The allostatic controller had higher cost and slightly higher variation than its matched ungated ablation, although both moved much less than naive and periodic policies.}]{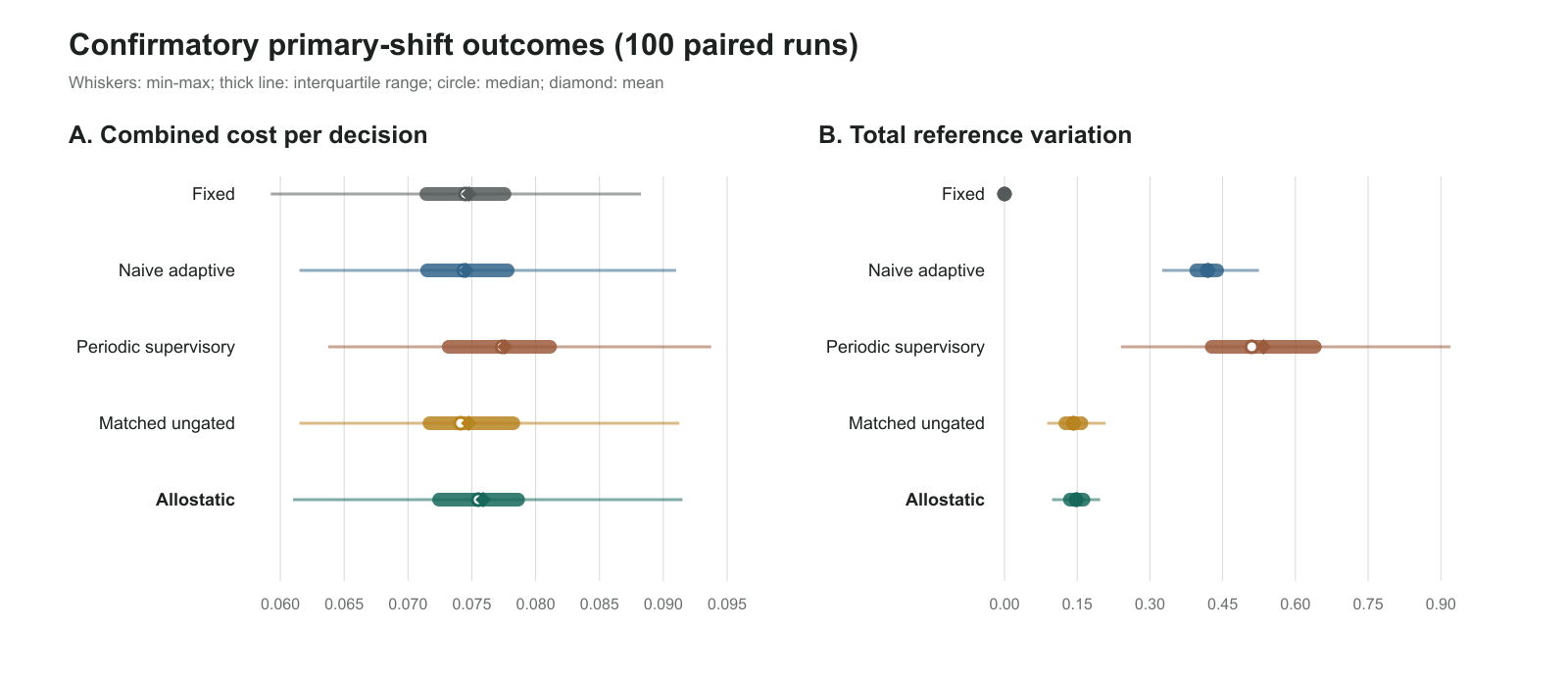}
\caption{Distribution summaries for decision cost and total reference
variation across the 100 paired confirmatory runs. The allostatic
controller had higher cost and slightly higher variation than its
matched ungated ablation, although both moved much less than naive and
periodic policies.}
\end{figure}

The selected fixed reference and naive adaptive policy also had lower
mean cost than the allostatic policy. Descriptive paired sign-flip tests
on the primary shift found allostatic cost higher than fixed, naive, and
matched policies after Holm adjustment (\(p<0.001\) for each; Appendix
B). At the run level, allostatic cost was lower than matched cost in 24
runs, equal in five, and higher in 71. Paired costs were strongly
correlated (\(r=0.9461\)), indicating that the mean disadvantage was not
created by a small set of incomparable streams.

\subsection{Stationary safety and periodic
comparison}\label{stationary-safety-and-periodic-comparison}

H3 was a registered component of the intersection-union central claim
and met its component gates. Relative allostatic cost excess was 0.00577
versus fixed, with a one-sided 95\% upper bound of 0.00891, and -0.03862
versus periodic supervision, with upper bound -0.03286. Median maximum
drift was 0.01984 with upper bound 0.02176, below the 0.03 gate. Because
the central claim required both H1 and H3, H1 failure precludes a
central confirmatory success; H3 is not promoted into a stand-alone
claim.

The registered H2 component calculations also met their numerical gates:
allostatic cost excess versus periodic supervision was -0.02191 (upper
bound -0.01482), and reference-variation reduction was 0.72003 (lower
bound 0.70466). The frozen hierarchy required H1 and H3 to pass before
H2 could pass, so the refinement claim is not supported. The descriptive
conclusion is narrower: the bounded mature-evidence policy moved much
less than periodic empirical-risk resetting, but not less than its
matched ungated comparator, and the smoother behavior did not produce
lower primary cost.

\subsection{Evidence latency}\label{evidence-latency}

Adaptation delay is consistent with a candidate mechanism behind H1.
After the prevalence changed from 0.35 to 0.15 at decision 1,000, the
allostatic reference failed to enter the new oracle neighborhood within
the segment in 94 of 100 runs; mean delay, with nonadaptation coded as
the full segment, was 985 decisions. After the change from 0.15 to 0.65
at decision 2,000, 88 runs did not adapt within the segment and mean
delay was 967.5. At the return to 0.35, only five runs failed to adapt
and mean delay was 180.5.

The matched policy was also slow under the selected deadbanded law, but
less so: nonadaptation counts were 81, 77, and 6. Within this matched
construction, the evidence gates did not convert delayed labels into a
better-timed reference; they withheld information available to the
ungated comparator. The bundle comparison does not separately identify
the contribution of each gate, but its observed behavior is consistent
with mature evidence becoming evidence latency.

\begin{figure}
\centering
\includegraphics[width=1\linewidth,height=\textheight,keepaspectratio,alt={Mean reference trajectories across primary confirmatory runs. The allostatic trajectory moved smoothly but lagged the generator-derived oracle after the first two changes. The oracle was unavailable to every controller.}]{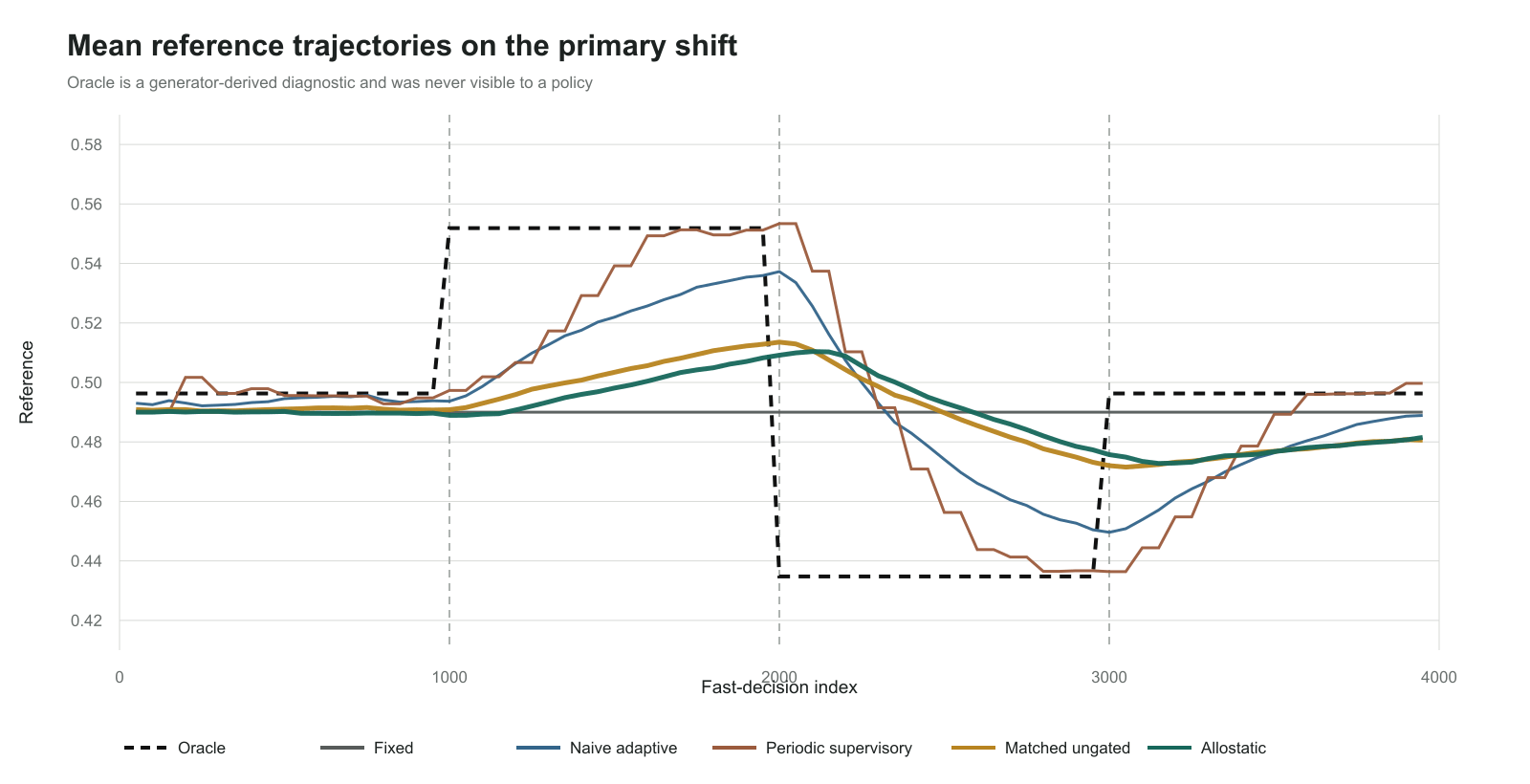}
\caption{Mean reference trajectories across primary confirmatory runs.
The allostatic trajectory moved smoothly but lagged the
generator-derived oracle after the first two changes. The oracle was
unavailable to every controller.}
\end{figure}

\subsection{Failure-boundary and no-signal
controls}\label{failure-boundary-and-no-signal-controls}

Under rapid alternation, allostatic mean cost was 0.60\% lower than
matched ungated adaptation, but remained worse than fixed. This
descriptive sign reversal does not rescue H1; it is consistent with
dependence on the relation among reveal delay, checkpoint interval, and
regime duration, which this experiment did not decompose causally.

The independent-label null was more damaging. Scores contained no label
information, yet all adaptive policies moved. Mean cost was 0.68203 for
fixed, 0.68797 for matched, and 0.69581 for allostatic control. In this
descriptive null comparison, allostatic cost exceeded matched by
1.139\%, and its mean total variation was 0.34938. Maturity alone did
not protect the controller from acting on uninformative evidence.

\subsection{Registered sensitivity
analysis}\label{registered-sensitivity-analysis}

Seven preregistered sensitivity cells varied reveal probability,
false-negative cost, and the minimum resolved fraction without retuning.
Lower reveal probability and stricter maturity made the allostatic
policy approach fixed behavior. At reveal probability 0.60, mean total
variation fell to 0.01731 while 65.35 cohorts became stale on average.
At minimum resolved fraction 0.90, total variation fell to 0.00110,
stale retirements rose to 71.45, and the policy held at 78.55 of 79
checkpoints. This was not evidence of successful conservatism: it was
near cessation of adaptation.

At reveal probability 0.90, allostatic variation increased to 0.16702
and mean cost remained above fixed and matched. Under false-negative
costs 1 and 4, adaptive policies could beat the fixed comparator in some
cells, but matched ungated adaptation still beat allostatic control. No
sensitivity cell replaces the confirmatory estimate.

\begin{figure}
\centering
\includegraphics[width=1\linewidth,height=\textheight,keepaspectratio,alt={Registered sensitivity analysis. Most cells retained a cost disadvantage versus the matched policy. Stricter maturity and lower reveal probability suppressed movement primarily by increasing stale-cohort retirement.}]{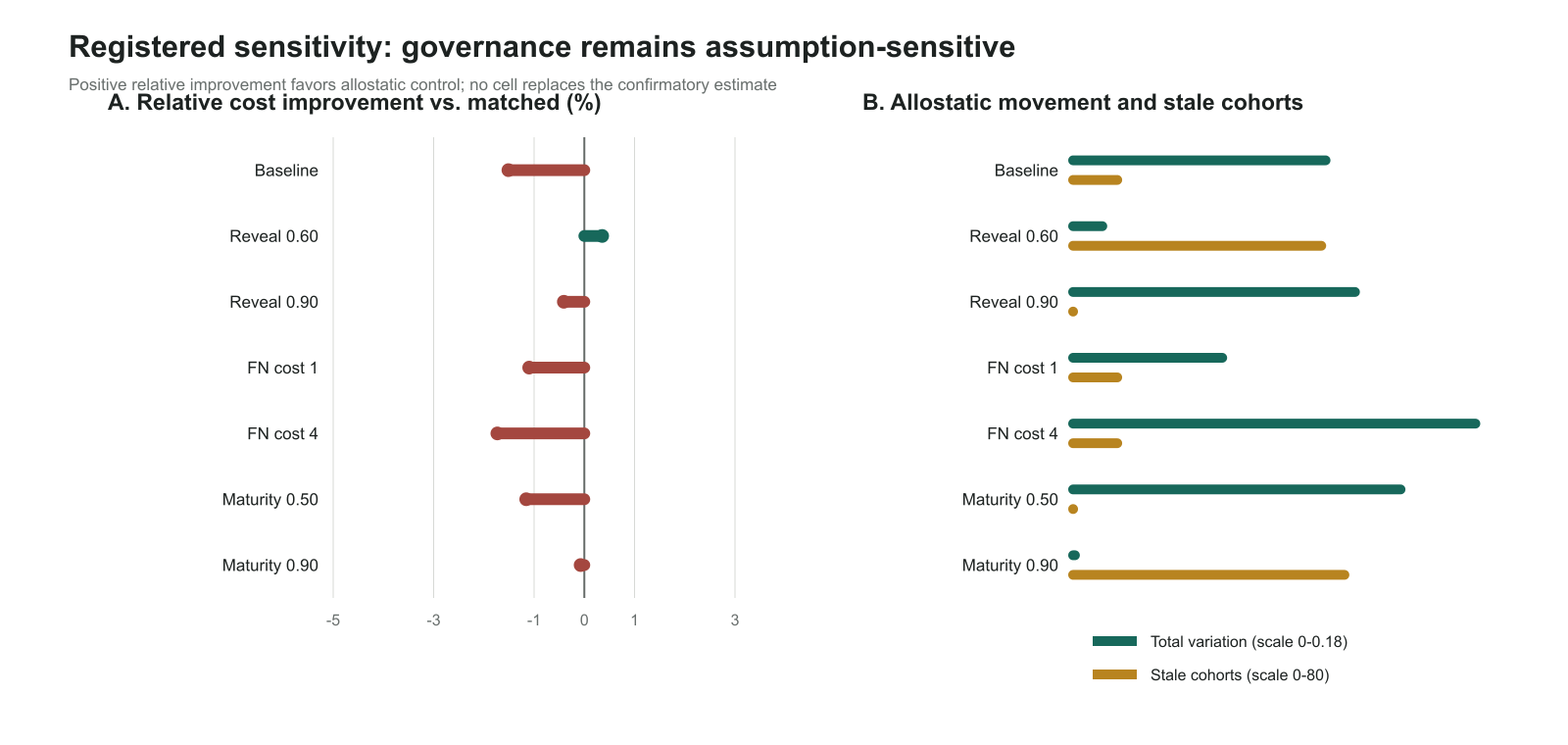}
\caption{Registered sensitivity analysis. Most cells retained a cost
disadvantage versus the matched policy. Stricter maturity and lower
reveal probability suppressed movement primarily by increasing
stale-cohort retirement.}
\end{figure}

\section{Interpretation}\label{interpretation}

The experiment rejects the paper's strongest proposed mechanism in the
declared regime. Evidence maturity, staleness, ordering, and
one-consumption rules did not improve a bounded adaptive reference.
Their cost was not merely lack of benefit: the confirmatory interval
supported modest harm relative to a matched ungated law. This result
falsifies the registered bundle; it does not identify maturity gating
alone as the cause.

The generator did not have a flat cost surface. Analytic integration
under the declared beta laws gives expected mean unconstrained
Bayes-placement cost 0.067515 and expected fixed-reference cost
0.074799. The observed fixed mean, 0.074785, closely matches that
calculation. The unconstrained difference relative to matched ungated is
9.69\%, while the zero-lag clairvoyant tracker constrained by the
registered checkpoint schedule and step cap achieves 5.02\%. At 50
decisions of imposed lag the diagnostic margin is 3.33\%; at 100 it is
1.63\%. Better reference placement was therefore available, and one
capped trajectory with immediate regime knowledge could clear the
practical gate. Whether any realizable policy inferring change from
delayed evidence could do so is not established.

The maturity mechanism was mismatched to the evidence clock. With
incomplete reveal, geometric delay, 50-decision cohorts, a 70\%
resolved-fraction threshold, and 300-decision stale limit, many cohorts
became usable only after their regime information had depreciated. The
oldest-first and one-consumption rules privileged completeness and
chronological processing over freshness. That is a defensible governance
choice, but in a non-stationary environment it can create avoidable lag.
A newest-mature-first arm, holding every other mechanism fixed, is the
direct post hoc diagnostic needed to distinguish an ordering failure
from a broader failure of maturity gating. It was not part of the
registered comparison and is not inferred here.

The frozen generator makes this pressure calculable. Let \(q=0.75\) be
reveal probability, \(\rho=0.02\) the geometric-delay parameter, and let
\(k\) be the number of decisions elapsed after the end of a 50-decision
cohort. Before the 600-decision cap becomes relevant, the expected
resolved count at that checkpoint is

\[
\mathbb{E}[M(k)] = q\sum_{d=k+1}^{k+50}\left[1-(1-\rho)^d\right].
\]

At \(k=0,50,100,\) and \(150\), the expected counts are 14.13, 28.99,
34.40, and 36.37. At integer resolution, the expectation first reaches
35 at \(k=111\), or 161 decisions after the cohort's first observation.
The controller evaluates evidence only at 50-decision checkpoints, so
the first registered checkpoint whose expected count reaches the
maturity threshold is \(k=150\), or 200 decisions after the cohort's
first observation. A stronger limit comes from incomplete reveal alone.
The eventual revealable count is \(R\sim\mathrm{Binomial}(50,0.75)\), so
\(\Pr(R\ge35)=0.8369\): even with unlimited waiting, 16.31\% of cohorts
cannot meet the threshold. Finite delay and staleness can only increase
that attrition. This calculation accounts for the order of the observed
stale-retirement burden, but not the full 985- and 967.5-decision
adaptation delays. Those longer delays are consistent with the joint
contribution of eligibility, oldest-first consumption, deadbanded local
movement, and distance to the new oracle neighborhood; their separate
effects were not identified by the registered contrast.

The result changes the status of several conceptual claims.

\begin{itemize}
\tightlist
\item
  Treating reference movement as a separate governance object remains
  coherent and useful.
\item
  Evidence time remains a useful diagnostic vocabulary, but ``wait for
  mature evidence'' is not a generally supported update prescription.
\item
  Scarce, cost-bearing evidence may be preferable to raw frequency as a
  design principle, but maturity and scarcity do not guarantee timely or
  informative evidence.
\item
  The matched ablation was necessary for interpreting this experiment.
  Comparisons only with fixed or periodic control would have made the
  tested policy look more favorable while failing to isolate its
  distinctive mechanism.
\item
  The generator had a non-flat reference-cost surface, and one capped
  immediate-detection trajectory cleared the 3\% gate; reachability for
  a realizable delayed-evidence policy remains unestablished.
\item
  In this environment, the tested mature-evidence specialization should
  be described as a falsifiable research hypothesis, not a validated
  design method. The result does not negate other allostatic-control
  architectures.
\end{itemize}

\subsection{Derived successor hypothesis: evidence-to-effect
catch-up}\label{derived-successor-hypothesis-evidence-to-effect-catch-up}

The negative result suggests a stricter question than whether evidence
has matured: can the complete evidence-to-effect pathway produce
material slow-loop correction before fast-loop operation accumulates an
unacceptable serviceability loss? That pathway includes outcome reveal,
eligibility, slow-loop use, bounded movement, and downstream effect. A
successor controller should therefore test whether bounded correction is
closing a retained, serviceability-relevant reference gap faster than
continued operation is enlarging it. If the gap cannot be closed in
time, ordinary adaptation should not continue merely because each
individual update remains admissible.

This catch-up criterion is a post-result hypothesis, not a conclusion
tested here. So are information-age weighting, explicit
value-of-information tests, change-point evidence, anchor-relative
replay, and recovery behavior when ordinary movement cannot catch up.
They require a new versioned model and separate preregistration. These
proposals cannot be used to reinterpret or replace the present outcome.

\section{Domain Illustrations and Research
Program}\label{domain-illustrations-and-research-program}

These illustrations are conceptual only. The synthetic negative result
neither establishes nor refutes efficacy in a real domain; it identifies
a candidate mechanism and failure condition that a domain-specific study
would have to test. In clinical alarm management, the defended object
could be an alarm boundary, the fast loop could evaluate current
measurements, and delayed clinically adjudicated outcomes plus response
burden could provide evidence. Alarm fatigue is well documented, but
fewer alarms are not by themselves a clinical benefit (Cvach 2012;
Sendelbach and Funk 2013; Lewandowska et al. 2020; Yang et al. 2025). A
credible study would require prospective oversight, safety envelopes,
established patient-specific baselines, and outcomes fixed before
evaluation.

Cloud autoscaling provides an engineering illustration. A service
objective or reserve posture could be the defended reference, local
allocation the fast action, and delayed degradation, queueing, fallback
cost, or overprovisioning the evidence (Hellerstein et al. 2004;
Podolskiy et al. 2018; Jeong and Jeong 2025). The present calculation
suggests a testable concern rather than a domain conclusion: if evidence
eligibility takes a substantial fraction of a workload regime's dwell
time, a maturity gate may become stale before it becomes useful.
Evaluation would require realistic traces, failure injection,
established autoscaling baselines, and a preregistered cost model.

\subsection{Future direction: allostatic
airframes}\label{future-direction-allostatic-airframes}

Future aircraft, autonomous platforms, and human-machine mission systems
may need governance of operating posture rather than fixed thresholds
alone. The defended object could include an autonomy posture, workload
boundary, reserve posture, or mission objective whose appropriateness
changes with vehicle condition, operator state, mission demand, and the
environment. This question is adjacent to established work on levels of
automation and adaptable human-automation teaming (Parasuraman et al.
2000; Calhoun 2021).

This paper does not propose or validate a flight-control law. The fast
loop in such a system could use established flight-control and
allocation methods, while a separately governed layer asks whether the
posture those methods serve remains appropriate. Any credible test would
require hard safety envelopes, human override, high-fidelity simulation,
failure injection, and predeclared criteria that prevent lower workload
or higher mission performance from being purchased through unacceptable
loss of vehicle or human safety.

The three illustrations can therefore be used as a compact design map:

{\def\LTcaptype{none} % do not increment counter
\begin{longtable}[]{@{}
  >{\raggedright\arraybackslash}p{(\linewidth - 6\tabcolsep) * \real{0.2500}}
  >{\raggedright\arraybackslash}p{(\linewidth - 6\tabcolsep) * \real{0.2500}}
  >{\raggedright\arraybackslash}p{(\linewidth - 6\tabcolsep) * \real{0.2500}}
  >{\raggedright\arraybackslash}p{(\linewidth - 6\tabcolsep) * \real{0.2500}}@{}}
\toprule\noalign{}
\begin{minipage}[b]{\linewidth}\raggedright
Domain
\end{minipage} & \begin{minipage}[b]{\linewidth}\raggedright
Governed goal or reference
\end{minipage} & \begin{minipage}[b]{\linewidth}\raggedright
Cost-bearing serviceability evidence
\end{minipage} & \begin{minipage}[b]{\linewidth}\raggedright
Required falsification or safety boundary
\end{minipage} \\
\midrule\noalign{}
\endhead
\bottomrule\noalign{}
\endlastfoot
Clinical alarms & Alarm threshold or alarm posture & Clinician response
burden, missed or delayed critical events, intervention timing, and
patient outcome & Burden must not be reduced by increasing missed
deterioration or delaying intervention. \\
Cloud autoscaling & Service objective, reserve, or resource posture &
Queueing, fallback, recovery delay, service violations, unmet demand,
and avoidable overprovisioning & Must beat established reactive,
predictive, or managed baselines on predeclared cost and service
measures under held-out traces. \\
Future airframes and human-machine systems & Autonomy, workload,
reserve, or mission posture & Vehicle integrity, operator workload and
recovery, mission demand, and environmental constraint & No deployment
claim without hard envelopes, human override, high-fidelity testing, and
domain safety evidence. \\
\end{longtable}
}

The architecture transfers across these domains; the evidence and safety
case do not. Each domain must define what the system is trying to
protect, how the current goal can fail to serve it, and what result
would kill the proposed mechanism.

\section{Limitations and Safety}\label{limitations-and-safety}

The experiment is intentionally narrow. It contains no plant dynamics,
state estimator, actuator, disturbance model, or closed-loop stability
analysis. Its reference is scalar; its regime changes and score family
are generated; its labels are delayed but ultimately scored under a
known synthetic law; and its policies operate on a common-information
replay contract. The results therefore concern a decision-reference
experiment, not general control-system performance.

Development selection used the same two scenario classes later evaluated
in fresh streams, which protects confirmatory independence at the
random-stream level but not against structural alignment to the
generator family. The 100 paired runs support the registered simulation
estimands but do not establish real-world effect sizes. Oracle
neighborhoods are analytical conveniences of the synthetic generator.
They are not available in deployed domains. The imposed-lag clairvoyant
tracker is a greedy deterministic diagnostic, not a globally optimal
controller or a universal upper bound. Its lag values were imposed
rather than estimated from the tested policies, so falling below 3\% at
100 decisions does not prove that every policy with delayed evidence
must fail.

The stationary gate bounded observed drift and relative cost under one
stationary generator; it is not a safety proof. Projection and step caps
bound one update but do not establish trajectory stability, constraint
satisfaction under plant dynamics, or resilience to adversarial
evidence. The no-signal control shows that mature evidence can still be
uninformative.

The mechanism evaluated here is a synthetic research construction, not a
production control architecture. Generalization to deployed systems
requires domain-specific state, actuation, evidence, cost, safety, and
governance models.

\section{Conclusion}\label{conclusion}

The motivating question survives: when should a system stop defending
its current reference? The first tested answer does not. In a
preregistered environment with delayed and incomplete outcomes,
mature-cohort allostatic governance was 1.51\% worse than mechanically
matched ungated adaptation. The registered lower-bound gate and the
corresponding upper summary were both on the side of harm. The policy
stayed within its stationary gates and moved less than periodic
supervision, but those secondary properties did not compensate for
evidence latency or establish the central claim.

The practical lesson is methodological. Reference movement is a distinct
and testable governance problem, but an evidence gate must earn its
delay, exclusions, and complexity. Changing environments do not
guarantee that moving references outperform a fixed compromise, and
mature evidence is not necessarily timely evidence. Future allostatic
mechanisms should be judged against matched adaptive ablations,
preregistered failure criteria, no-signal controls, and explicit
relations among evidence delay, regime duration, cost-surface curvature,
and bounded correction capacity. The next testable question is whether
the full evidence-to-effect pathway can close serviceability loss faster
than fast-loop operation accumulates it. Until such tests succeed, this
mature-evidence specialization is best treated as a disciplined,
falsifiable research program rather than an empirically validated
control method.

\section{Data, Code, and AI-Assistance
Statement}\label{data-code-and-ai-assistance-statement}

A checksum-bound replication package prepared for this paper contains
the frozen experiment contract, source, parameter-selection evidence,
raw confirmatory results, analysis code, and integrity receipts. Its
artifact record states the release scope and reproduces the hashes
reported here. The package also includes a frozen environment
specification and a verification command that recomputes artifact
digests before reproducing the analysis from raw results.

The author used large language models for literature discovery, drafting
assistance, software implementation, code review, and adversarial
manuscript review. Claims, experimental design, frozen hypotheses,
result interpretation, and final authorship responsibility remain with
the author. Citations were checked against source metadata rather than
accepted from model recall.

\section{Appendix A: Registered Metrics and
Gates}\label{appendix-a-registered-metrics-and-gates}

For run \(i\) with \(T=4000\) fast decisions, the normalized
false-instantiation burden was \(T^{-1}\sum_t\mathbf{1}\{a_t=1,y_t=0\}\)
and the normalized suppressed-distinction burden was
\(2T^{-1}\sum_t\mathbf{1}\{a_t=0,y_t=1\}\). Their sum was the registered
combined cost per decision. Total reference variation was
\(\sum_j|r_{j+1}-r_j|\); maximum drift was \(\max_j|r_j-r_0|\); and a
reversal was a change in the nonzero update direction. Holds, stale
retirements, and boundary contacts were counted directly from the
slow-loop record.

Adaptation delay after a registered regime boundary was the first
checkpoint at which the reference entered the new oracle neighborhood
and met the registered persistence rule. Failure within the
1,000-decision segment was retained as nonadaptation and assigned delay
1,000. This convention penalizes rather than censors failure to adapt.

{\def\LTcaptype{none} % do not increment counter
\begin{longtable}[]{@{}
  >{\raggedright\arraybackslash}p{(\linewidth - 6\tabcolsep) * \real{0.2308}}
  >{\raggedright\arraybackslash}p{(\linewidth - 6\tabcolsep) * \real{0.2308}}
  >{\raggedleft\arraybackslash}p{(\linewidth - 6\tabcolsep) * \real{0.3077}}
  >{\raggedright\arraybackslash}p{(\linewidth - 6\tabcolsep) * \real{0.2308}}@{}}
\toprule\noalign{}
\begin{minipage}[b]{\linewidth}\raggedright
Registered test
\end{minipage} & \begin{minipage}[b]{\linewidth}\raggedright
Estimand and gate
\end{minipage} & \begin{minipage}[b]{\linewidth}\raggedleft
Observed
\end{minipage} & \begin{minipage}[b]{\linewidth}\raggedright
Disposition
\end{minipage} \\
\midrule\noalign{}
\endhead
\bottomrule\noalign{}
\endlastfoot
H1 mechanism & Relative cost improvement vs.~matched; lower one-sided
95\% bound \(>0\) and point \(\ge0.03\) & point -0.01512; lower
-0.01906; upper -0.01141 & Fail \\
H3 vs.~fixed & Stationary relative cost excess; upper one-sided 95\%
bound \(<0.02\) & point 0.00577; upper 0.00891 & Component pass \\
H3 vs.~periodic & Stationary relative cost excess; upper one-sided 95\%
bound \(<0.02\) & point -0.03862; upper -0.03286 & Component pass \\
H3 drift & Stationary median maximum drift; upper one-sided 95\% bound
\(\le0.03\) & median 0.01984; upper 0.02176 & Component pass \\
H2 cost & Relative cost excess vs.~periodic; registered noninferiority
bound & point -0.02191; upper -0.01482 & Component pass only \\
H2 variation & Relative variation reduction vs.~periodic; registered
lower bound & point 0.72003; lower 0.70466 & Component pass only \\
\end{longtable}
}

H3 is reported because it was a registered component of the
intersection-union central claim; meeting H3 does not rescue the failed
H1. H2 component values are reported because they were registered, but
the conditional refinement could not pass after the hierarchy failed.
Neither set of component values establishes a secondary success claim.

\section{Appendix B: Full Scenario
Summaries}\label{appendix-b-full-scenario-summaries}

The table reports mean combined cost and mean total reference variation
across 100 runs in each frozen scenario. No result was excluded,
replaced, or selected for direction.

{\def\LTcaptype{none} % do not increment counter
\begin{longtable}[]{@{}
  >{\raggedright\arraybackslash}p{(\linewidth - 10\tabcolsep) * \real{0.1364}}
  >{\raggedright\arraybackslash}p{(\linewidth - 10\tabcolsep) * \real{0.1364}}
  >{\raggedleft\arraybackslash}p{(\linewidth - 10\tabcolsep) * \real{0.1818}}
  >{\raggedleft\arraybackslash}p{(\linewidth - 10\tabcolsep) * \real{0.1818}}
  >{\raggedleft\arraybackslash}p{(\linewidth - 10\tabcolsep) * \real{0.1818}}
  >{\raggedleft\arraybackslash}p{(\linewidth - 10\tabcolsep) * \real{0.1818}}@{}}
\toprule\noalign{}
\begin{minipage}[b]{\linewidth}\raggedright
Scenario
\end{minipage} & \begin{minipage}[b]{\linewidth}\raggedright
Policy
\end{minipage} & \begin{minipage}[b]{\linewidth}\raggedleft
Mean (SD) cost
\end{minipage} & \begin{minipage}[b]{\linewidth}\raggedleft
Median {[}IQR{]} cost
\end{minipage} & \begin{minipage}[b]{\linewidth}\raggedleft
Mean variation
\end{minipage} & \begin{minipage}[b]{\linewidth}\raggedleft
Mean max drift
\end{minipage} \\
\midrule\noalign{}
\endhead
\bottomrule\noalign{}
\endlastfoot
Slow shift & Fixed & 0.074785 (0.005112) & 0.074500 {[}0.006250{]} &
0.000000 & 0.000000 \\
Slow shift & Naive adaptive & 0.074558 (0.005207) & 0.074375
{[}0.006250{]} & 0.420496 & 0.059169 \\
Slow shift & Periodic supervisory & 0.077588 (0.005967) & 0.077375
{[}0.008000{]} & 0.534300 & 0.092500 \\
Slow shift & Matched ungated & 0.074758 (0.005335) & 0.074125
{[}0.006750{]} & 0.142655 & 0.032446 \\
Slow shift & Allostatic & 0.075888 (0.005322) & 0.075500 {[}0.006250{]}
& 0.149590 & 0.030047 \\
Stationary & Fixed & 0.073200 (0.004444) & 0.073125 {[}0.006000{]} &
0.000000 & 0.000000 \\
Stationary & Naive adaptive & 0.073903 (0.004532) & 0.074125
{[}0.006750{]} & 0.392151 & 0.034793 \\
Stationary & Periodic supervisory & 0.076580 (0.005438) & 0.076125
{[}0.008000{]} & 0.426900 & 0.062800 \\
Stationary & Matched ungated & 0.073630 (0.004460) & 0.073750
{[}0.006750{]} & 0.122702 & 0.022284 \\
Stationary & Allostatic & 0.073623 (0.004364) & 0.073875 {[}0.006500{]}
& 0.129064 & 0.020776 \\
Rapid alternation & Fixed & 0.074900 (0.004586) & 0.074625
{[}0.006250{]} & 0.000000 & 0.000000 \\
Rapid alternation & Naive adaptive & 0.076710 (0.004593) & 0.076500
{[}0.006000{]} & 0.423567 & 0.034009 \\
Rapid alternation & Periodic supervisory & 0.078605 (0.005299) &
0.078750 {[}0.007500{]} & 0.422200 & 0.056400 \\
Rapid alternation & Matched ungated & 0.075803 (0.004619) & 0.075750
{[}0.006750{]} & 0.138208 & 0.023149 \\
Rapid alternation & Allostatic & 0.075350 (0.004767) & 0.075250
{[}0.006500{]} & 0.167602 & 0.023800 \\
Independent-label null & Fixed & 0.682033 (0.011348) & 0.681375
{[}0.014750{]} & 0.000000 & 0.000000 \\
Independent-label null & Naive adaptive & 0.691542 (0.021438) & 0.689250
{[}0.029500{]} & 0.830944 & 0.203977 \\
Independent-label null & Periodic supervisory & 0.595480 (0.013628) &
0.596000 {[}0.019500{]} & 1.963000 & 0.259400 \\
Independent-label null & Matched ungated & 0.687973 (0.016444) &
0.687250 {[}0.024750{]} & 0.407301 & 0.108427 \\
Independent-label null & Allostatic & 0.695808 (0.017081) & 0.694500
{[}0.020750{]} & 0.349376 & 0.096028 \\
\end{longtable}
}

Mean holds and stale retirements are reported below as
\texttt{holds\ /\ stale} for all policies and scenarios. Each run had 79
slow checkpoints.

{\def\LTcaptype{none} % do not increment counter
\begin{longtable}[]{@{}
  >{\raggedright\arraybackslash}p{(\linewidth - 10\tabcolsep) * \real{0.1304}}
  >{\raggedleft\arraybackslash}p{(\linewidth - 10\tabcolsep) * \real{0.1739}}
  >{\raggedleft\arraybackslash}p{(\linewidth - 10\tabcolsep) * \real{0.1739}}
  >{\raggedleft\arraybackslash}p{(\linewidth - 10\tabcolsep) * \real{0.1739}}
  >{\raggedleft\arraybackslash}p{(\linewidth - 10\tabcolsep) * \real{0.1739}}
  >{\raggedleft\arraybackslash}p{(\linewidth - 10\tabcolsep) * \real{0.1739}}@{}}
\toprule\noalign{}
\begin{minipage}[b]{\linewidth}\raggedright
Scenario
\end{minipage} & \begin{minipage}[b]{\linewidth}\raggedleft
Fixed
\end{minipage} & \begin{minipage}[b]{\linewidth}\raggedleft
Naive
\end{minipage} & \begin{minipage}[b]{\linewidth}\raggedleft
Periodic
\end{minipage} & \begin{minipage}[b]{\linewidth}\raggedleft
Matched
\end{minipage} & \begin{minipage}[b]{\linewidth}\raggedleft
Allostatic
\end{minipage} \\
\midrule\noalign{}
\endhead
\bottomrule\noalign{}
\endlastfoot
Slow shift & 79.00 / 0.00 & 13.26 / 0.00 & 62.79 / 0.00 & 40.47 / 0.00 &
31.06 / 11.63 \\
Stationary & 79.00 / 0.00 & 13.97 / 0.00 & 64.98 / 0.00 & 43.84 / 0.00 &
32.25 / 12.36 \\
Rapid alternation & 79.00 / 0.00 & 13.66 / 0.00 & 63.84 / 0.00 & 41.32 /
0.00 & 30.54 / 12.36 \\
Independent null & 79.00 / 0.00 & 6.01 / 0.00 & 63.24 / 0.00 & 17.55 /
0.00 & 21.93 / 12.36 \\
\end{longtable}
}

The following descriptive sign-flip tests use the preregistered 10,000
permutations and Holm adjustment within the primary scenario's
combined-cost family. Differences are oriented as comparator minus
allostatic, so a negative value means lower cost for the comparator.
These tests support description of the policy ordering; they are not
replacements for H1.

{\def\LTcaptype{none} % do not increment counter
\begin{longtable}[]{@{}
  >{\raggedright\arraybackslash}p{(\linewidth - 6\tabcolsep) * \real{0.2000}}
  >{\raggedleft\arraybackslash}p{(\linewidth - 6\tabcolsep) * \real{0.2667}}
  >{\raggedleft\arraybackslash}p{(\linewidth - 6\tabcolsep) * \real{0.2667}}
  >{\raggedleft\arraybackslash}p{(\linewidth - 6\tabcolsep) * \real{0.2667}}@{}}
\toprule\noalign{}
\begin{minipage}[b]{\linewidth}\raggedright
Primary comparator
\end{minipage} & \begin{minipage}[b]{\linewidth}\raggedleft
Mean paired difference
\end{minipage} & \begin{minipage}[b]{\linewidth}\raggedleft
Raw \(p\)
\end{minipage} & \begin{minipage}[b]{\linewidth}\raggedleft
Holm-adjusted \(p\)
\end{minipage} \\
\midrule\noalign{}
\endhead
\bottomrule\noalign{}
\endlastfoot
Fixed & -0.0011025 & \(<0.0001\) & \(<0.001\) \\
Naive adaptive & -0.0013300 & \(<0.0001\) & \(<0.001\) \\
Periodic supervisory & 0.0017000 & \(<0.0001\) & \(<0.001\) \\
Matched ungated & -0.0011300 & \(<0.0001\) & \(<0.001\) \\
\end{longtable}
}

The periodic policy's lower mean cost in the independent-label null
reflects its movement toward a finite-sample empirical-risk minimizer
under the asymmetric label cost. It does not show that the score
contained information; the score was independent of the label by
construction. The null therefore tests policy behavior under
uninformative ranking, not the existence of a universally optimal fixed
action.

\subsection{Direction-specific and sensitivity
summaries}\label{direction-specific-and-sensitivity-summaries}

The primary segment table reports \texttt{mean;\ median\ {[}IQR{]}}
combined cost. Segments correspond to prevalence values 0.35, 0.15,
0.65, and 0.35. Direction-specific reporting prevents the aggregate
estimate from concealing asymmetric response to downward, upward, and
return shifts.

{\def\LTcaptype{none} % do not increment counter
\begin{longtable}[]{@{}
  >{\raggedright\arraybackslash}p{(\linewidth - 8\tabcolsep) * \real{0.1579}}
  >{\raggedleft\arraybackslash}p{(\linewidth - 8\tabcolsep) * \real{0.2105}}
  >{\raggedleft\arraybackslash}p{(\linewidth - 8\tabcolsep) * \real{0.2105}}
  >{\raggedleft\arraybackslash}p{(\linewidth - 8\tabcolsep) * \real{0.2105}}
  >{\raggedleft\arraybackslash}p{(\linewidth - 8\tabcolsep) * \real{0.2105}}@{}}
\toprule\noalign{}
\begin{minipage}[b]{\linewidth}\raggedright
Policy
\end{minipage} & \begin{minipage}[b]{\linewidth}\raggedleft
Segment 0
\end{minipage} & \begin{minipage}[b]{\linewidth}\raggedleft
Segment 1
\end{minipage} & \begin{minipage}[b]{\linewidth}\raggedleft
Segment 2
\end{minipage} & \begin{minipage}[b]{\linewidth}\raggedleft
Segment 3
\end{minipage} \\
\midrule\noalign{}
\endhead
\bottomrule\noalign{}
\endlastfoot
Fixed & 0.07467; 0.07450 {[}0.01000{]} & 0.06578; 0.06600 {[}0.01100{]}
& 0.08469; 0.08400 {[}0.01700{]} & 0.07400; 0.07400 {[}0.01200{]} \\
Naive adaptive & 0.07554; 0.07600 {[}0.01100{]} & 0.05689; 0.05550
{[}0.01200{]} & 0.08755; 0.08700 {[}0.01700{]} & 0.07825; 0.07700
{[}0.01500{]} \\
Periodic supervisory & 0.07829; 0.07800 {[}0.01300{]} & 0.05781; 0.05700
{[}0.01200{]} & 0.09286; 0.09300 {[}0.01800{]} & 0.08139; 0.08100
{[}0.01300{]} \\
Matched ungated & 0.07483; 0.07500 {[}0.01000{]} & 0.06117; 0.06000
{[}0.01200{]} & 0.08727; 0.08600 {[}0.01700{]} & 0.07576; 0.07550
{[}0.01600{]} \\
Allostatic & 0.07488; 0.07500 {[}0.01200{]} & 0.06320; 0.06300
{[}0.01200{]} & 0.08891; 0.08800 {[}0.01800{]} & 0.07656; 0.07500
{[}0.01300{]} \\
\end{longtable}
}

The complete registered one-factor sensitivity cost grid follows. Values
are mean combined cost over 100 runs; no cell was used to replace the
primary estimate.

{\def\LTcaptype{none} % do not increment counter
\begin{longtable}[]{@{}
  >{\raggedright\arraybackslash}p{(\linewidth - 10\tabcolsep) * \real{0.1304}}
  >{\raggedleft\arraybackslash}p{(\linewidth - 10\tabcolsep) * \real{0.1739}}
  >{\raggedleft\arraybackslash}p{(\linewidth - 10\tabcolsep) * \real{0.1739}}
  >{\raggedleft\arraybackslash}p{(\linewidth - 10\tabcolsep) * \real{0.1739}}
  >{\raggedleft\arraybackslash}p{(\linewidth - 10\tabcolsep) * \real{0.1739}}
  >{\raggedleft\arraybackslash}p{(\linewidth - 10\tabcolsep) * \real{0.1739}}@{}}
\toprule\noalign{}
\begin{minipage}[b]{\linewidth}\raggedright
Sensitivity cell
\end{minipage} & \begin{minipage}[b]{\linewidth}\raggedleft
Fixed
\end{minipage} & \begin{minipage}[b]{\linewidth}\raggedleft
Naive
\end{minipage} & \begin{minipage}[b]{\linewidth}\raggedleft
Periodic
\end{minipage} & \begin{minipage}[b]{\linewidth}\raggedleft
Matched
\end{minipage} & \begin{minipage}[b]{\linewidth}\raggedleft
Allostatic
\end{minipage} \\
\midrule\noalign{}
\endhead
\bottomrule\noalign{}
\endlastfoot
Baseline & 0.074785 & 0.074558 & 0.077588 & 0.074758 & 0.075888 \\
Reveal probability 0.60 & 0.074785 & 0.075080 & 0.078585 & 0.075278 &
0.075010 \\
Reveal probability 0.90 & 0.074785 & 0.074583 & 0.076768 & 0.075358 &
0.075665 \\
Suppressed-distinction cost 1 & 0.056798 & 0.052978 & 0.054820 &
0.054193 & 0.054788 \\
Suppressed-distinction cost 4 & 0.110760 & 0.103118 & 0.108455 &
0.104515 & 0.106325 \\
Minimum resolved fraction 0.50 & 0.074785 & 0.074558 & 0.077588 &
0.074758 & 0.075623 \\
Minimum resolved fraction 0.90 & 0.074785 & 0.074558 & 0.077588 &
0.074758 & 0.074813 \\
\end{longtable}
}

Allostatic movement and eligibility diagnostics for the same cells are:

{\def\LTcaptype{none} % do not increment counter
\begin{longtable}[]{@{}
  >{\raggedright\arraybackslash}p{(\linewidth - 6\tabcolsep) * \real{0.2000}}
  >{\raggedleft\arraybackslash}p{(\linewidth - 6\tabcolsep) * \real{0.2667}}
  >{\raggedleft\arraybackslash}p{(\linewidth - 6\tabcolsep) * \real{0.2667}}
  >{\raggedleft\arraybackslash}p{(\linewidth - 6\tabcolsep) * \real{0.2667}}@{}}
\toprule\noalign{}
\begin{minipage}[b]{\linewidth}\raggedright
Sensitivity cell
\end{minipage} & \begin{minipage}[b]{\linewidth}\raggedleft
Mean variation
\end{minipage} & \begin{minipage}[b]{\linewidth}\raggedleft
Mean holds
\end{minipage} & \begin{minipage}[b]{\linewidth}\raggedleft
Mean stale retirements
\end{minipage} \\
\midrule\noalign{}
\endhead
\bottomrule\noalign{}
\endlastfoot
Baseline & 0.149590 & 31.06 & 11.63 \\
Reveal probability 0.60 & 0.017315 & 73.76 & 65.35 \\
Reveal probability 0.90 & 0.167017 & 18.74 & 0.00 \\
Suppressed-distinction cost 1 & 0.088362 & 36.56 & 11.63 \\
Suppressed-distinction cost 4 & 0.238356 & 26.36 & 11.63 \\
Minimum resolved fraction 0.50 & 0.194009 & 22.42 & 0.01 \\
Minimum resolved fraction 0.90 & 0.001096 & 78.55 & 71.45 \\
\end{longtable}
}

\section{Appendix C: Reproducibility
Manifest}\label{appendix-c-reproducibility-manifest}

{\def\LTcaptype{none} % do not increment counter
\begin{longtable}[]{@{}
  >{\raggedright\arraybackslash}p{(\linewidth - 2\tabcolsep) * \real{0.5000}}
  >{\raggedright\arraybackslash}p{(\linewidth - 2\tabcolsep) * \real{0.5000}}@{}}
\toprule\noalign{}
\begin{minipage}[b]{\linewidth}\raggedright
Artifact
\end{minipage} & \begin{minipage}[b]{\linewidth}\raggedright
SHA-256 or receipt identity
\end{minipage} \\
\midrule\noalign{}
\endhead
\bottomrule\noalign{}
\endlastfoot
Frozen source commit &
\seqsplit{7fd2d30e970612127cff2ebea7b860e399b6b7f5} \\
Frozen source tree &
\seqsplit{e5d3a2a33d8ff3aed3980070828f5879481bb9bad0d7a1ab5301f3011de612ec} \\
Runtime lock &
\seqsplit{729f321758632063c0af996551edefa8fbfecdf8db82f5a680de18fe459080eb} \\
Selected parameters &
\seqsplit{b764f6fd32e7708e13b71394d4dfad4202cd66d8398813853bf4f8a815e18c50} \\
Frozen preregistration &
\seqsplit{b66740c8cc986bf9876e3e10ada043ec5e390bc855fbb6c2614a1709e53008d0} \\
Execution manifest &
\seqsplit{6e951ed15b8ad60890ec57c528bf8aca32b0f5846b564a164141723e20432648} \\
Raw-result custody receipt &
\seqsplit{af92fad03846a64dc0368363552504cd9d8a0640cd60ec69b98122704651784c} \\
Analysis report and exact replay &
\seqsplit{0e103002f718ba3e6a8eb4103369e27add629346b323a19086ab3f862a67dd92} \\
\end{longtable}
}

The preregistration receipt binds the frozen commit and source-tree
digest. The raw execution manifest binds separate primary-result,
sensitivity-result, and telemetry objects. Object-level retention
extends through July 16, 2027. The replication package provides a
verification command that recomputes these digests before reproducing
the analysis.

\section*{References}\label{bibliography}
\addcontentsline{toc}{section}{References}

\protect\phantomsection\label{refs}


\begin{CSLReferences}{1}{1}
\bibitem[\citeproctext]{ref-amrit2011_economic_mpc}
Amrit, Rishi, James B. Rawlings, and David Angeli. 2011. {``Economic
Optimization Using Model Predictive Control with a Terminal Cost.''}
\emph{Annual Reviews in Control} 35 (2): 178--86.
\url{https://doi.org/10.1016/j.arcontrol.2011.10.011}.

\bibitem[\citeproctext]{ref-ashby1960_design_brain}
Ashby, W. Ross. 1960. \emph{Design for a Brain: The Origin of Adaptive
Behaviour}. 2nd ed. Chapman \& Hall.
\url{https://www.ashby.info/Ashby\%20-\%20Design\%20for\%20a\%20Brain\%20-\%20The\%20Origin\%20of\%20Adaptive\%20Behavior.pdf}.

\bibitem[\citeproctext]{ref-astrom_murray_feedback_systems}
Astrom, Karl J., and Richard M. Murray. 2008. \emph{Feedback Systems: An
Introduction for Scientists and Engineers}. Princeton University Press.
\url{https://fbswiki.org/wiki/index.php/Feedback_Systems:_An_Introduction_for_Scientists_and_Engineers}.

\bibitem[\citeproctext]{ref-astrom_wittenmark_adaptive_control}
Astrom, Karl J., and Bjorn Wittenmark. 1995. \emph{Adaptive Control}.
2nd ed. Addison-Wesley.
\url{https://portal.research.lu.se/en/publications/adaptive-control-2-ed/}.

\bibitem[\citeproctext]{ref-borkar1997_two_timescales}
Borkar, Vivek S. 1997. {``Stochastic Approximation with Two Time
Scales.''} \emph{Systems \& Control Letters} 29 (5): 291--94.
\url{https://doi.org/10.1016/s0167-6911(97)90015-3}.

\bibitem[\citeproctext]{ref-calhoun2021_adaptable_automation}
Calhoun, Gloria. 2021. {``Adaptable (Not Adaptive) Automation: Forefront
of Human-Automation Teaming.''} \emph{Human Factors} 64 (2): 269--77.
\url{https://doi.org/10.1177/00187208211037457}.

\bibitem[\citeproctext]{ref-cvach2012_alarm_fatigue}
Cvach, Maria. 2012. {``Monitor Alarm Fatigue: An Integrative Review.''}
\emph{Biomedical Instrumentation \& Technology} 46 (4): 268--77.
\url{https://doi.org/10.2345/0899-8205-46.4.268}.

\bibitem[\citeproctext]{ref-gama2014_concept_drift}
Gama, Joao, Indre Zliobaite, Albert Bifet, Mykola Pechenizkiy, and
Abdelhamid Bouchachia. 2014. {``A Survey on Concept Drift Adaptation.''}
\emph{ACM Computing Surveys} 46 (4): 1--37.
\url{https://doi.org/10.1145/2523813}.

\bibitem[\citeproctext]{ref-garone2017_reference_governors}
Garone, Emanuele, Stefano Di Cairano, and Ilya Kolmanovsky. 2017.
{``Reference and Command Governors for Systems with Constraints: A
Survey on Theory and Applications.''} \emph{Automatica} 75: 306--28.
\url{https://doi.org/10.1016/j.automatica.2016.08.013}.

\bibitem[\citeproctext]{ref-hellerstein2004_feedback_computing}
Hellerstein, Joseph L., Yixin Diao, Sujay Parekh, and Dawn M. Tilbury.
2004. \emph{Feedback Control of Computing Systems}. Wiley/IEEE Press.
\url{https://doi.org/10.1002/047166880X}.

\bibitem[\citeproctext]{ref-hewing2020_learning_based_mpc}
Hewing, Lukas, Kim P. Wabersich, Marcel Menner, and Melanie N.
Zeilinger. 2020. {``Learning-Based Model Predictive Control: Toward Safe
Learning in Control.''} \emph{Annual Review of Control, Robotics, and
Autonomous Systems} 3 (1): 269--96.
\url{https://doi.org/10.1146/annurev-control-090419-075625}.

\bibitem[\citeproctext]{ref-jeong2025_autoscaling_survey}
Jeong, Byeonghui, and Young-Sik Jeong. 2025. {``Autoscaling Techniques
in Cloud-Native Computing: A Comprehensive Survey.''} \emph{Computer
Science Review} 58: 100791.
\url{https://doi.org/10.1016/j.cosrev.2025.100791}.

\bibitem[\citeproctext]{ref-lewandowska2020_alarm_fatigue}
Lewandowska, Katarzyna, Magdalena Weisbrot, Aleksandra Cieloszyk,
Wioletta Medrzycka-Dabrowska, Sabina Krupa, and Dorota Ozga. 2020.
{``Impact of Alarm Fatigue on the Work of Nurses in an Intensive Care
Environment: A Systematic Review.''} \emph{International Journal of
Environmental Research and Public Health} 17 (22): 8409.
\url{https://doi.org/10.3390/ijerph17228409}.

\bibitem[\citeproctext]{ref-mayne2000_constrained_mpc}
Mayne, D. Q., J. B. Rawlings, C. V. Rao, and P. O. M. Scokaert. 2000.
{``Constrained Model Predictive Control: Stability and Optimality.''}
\emph{Automatica} 36 (6): 789--814.
\url{https://doi.org/10.1016/s0005-1098(99)00214-9}.

\bibitem[\citeproctext]{ref-parasuraman2000_automation_levels}
Parasuraman, Raja, Thomas B. Sheridan, and Christopher D. Wickens. 2000.
{``A Model for Types and Levels of Human Interaction with Automation.''}
\emph{IEEE Transactions on Systems, Man, and Cybernetics - Part A:
Systems and Humans} 30 (3): 286--97.
\url{https://doi.org/10.1109/3468.844354}.

\bibitem[\citeproctext]{ref-podolskiy2018_reactive_autoscaling}
Podolskiy, Vladimir, Anshul Jindal, and Michael Gerndt. 2018. {``IaaS
Reactive Autoscaling Performance Challenges.''} \emph{2018 IEEE 11th
International Conference on Cloud Computing}, 954--57.
\url{https://doi.org/10.1109/cloud.2018.00144}.

\bibitem[\citeproctext]{ref-ramadge_wonham_supervisory}
Ramadge, P. J., and W. M. Wonham. 1987. {``Supervisory Control of
Discrete Event Processes.''} In \emph{Lecture Notes in Control and
Information Sciences}. \url{https://doi.org/10.1007/bfb0006830}.

\bibitem[\citeproctext]{ref-rawlings2012_economic_mpc}
Rawlings, James B., David Angeli, and Cuyler N. Bates. 2012.
{``Fundamentals of Economic Model Predictive Control.''} \emph{2012 IEEE
51st IEEE Conference on Decision and Control}, 3851--61.
\url{https://doi.org/10.1109/cdc.2012.6425822}.

\bibitem[\citeproctext]{ref-rawlings_mayne_diehl_mpc}
Rawlings, James B., David Q. Mayne, and Moritz M. Diehl. 2026.
\emph{Model Predictive Control: Theory, Computation, and Design}. 2nd
ed. Nob Hill Publishing.
\url{https://sites.engineering.ucsb.edu/~jbraw/mpc/}.

\bibitem[\citeproctext]{ref-sanchez_fibla2010_allostatic}
Sanchez-Fibla, Marti, Ulysses Bernardet, Erez Wasserman, et al. 2010.
{``Allostatic Control for Robot Behavior Regulation: A Comparative
Rodent-Robot Study.''} \emph{Advances in Complex Systems} 13 (3):
377--403. \url{https://doi.org/10.1142/S0219525910002621}.

\bibitem[\citeproctext]{ref-schulkin_sterling2019_allostasis}
Schulkin, Jay, and Peter Sterling. 2019. {``Allostasis: A
Brain-Centered, Predictive Mode of Physiological Regulation.''}
\emph{Trends in Neurosciences} 42 (10): 740--52.
\url{https://doi.org/10.1016/j.tins.2019.07.010}.

\bibitem[\citeproctext]{ref-sendelbach_funk2013_alarm_fatigue}
Sendelbach, Sue, and Marjorie Funk. 2013. {``Alarm Fatigue: A Patient
Safety Concern.''} \emph{AACN Advanced Critical Care} 24 (4): 378--86.
\url{https://doi.org/10.4037/NCI.0b013e3182a903f9}.

\bibitem[\citeproctext]{ref-sterling2012_allostasis}
Sterling, Peter. 2012. {``Allostasis: A Model of Predictive
Regulation.''} \emph{Physiology \& Behavior} 106 (1): 5--15.
\url{https://doi.org/10.1016/j.physbeh.2011.06.004}.

\bibitem[\citeproctext]{ref-yang2025_alarm_management}
Yang, Chunqing, Ping Gao, Hanlin Ji, et al. 2025. {``Clinical Alarm
Management in Intensive Care Units: A Scoping Review.''} \emph{Nursing
in Critical Care} 30 (3): e70042.
\url{https://doi.org/10.1111/nicc.70042}.

\end{CSLReferences}
\end{document}